\documentclass[a4paper]{article}
\usepackage{spconf}
\usepackage{graphicx}
\usepackage{hyperref}
\usepackage{multirow}
\usepackage{array}
\usepackage{epstopdf}
\usepackage{algorithm}
\usepackage{algorithmic}
\usepackage{mathrsfs}
\usepackage{bm}
\usepackage{verbatim}
\usepackage{bm,url,amsmath,amssymb,subfigure,epsfig}

\title{Acoustics-guided evaluation (AGE): a new measure for estimating performance of speech enhancement algorithms for robust ASR}
\name{Li Chai$^1$, Jun Du$^2$, and Chin-Hui Lee$^3$}
\address{$^1$School of Data Science, University of Science and technology of China, Hefei, Anhui, P. R. China\\
  $^2$University of Science and technology of China, Hefei, Anhui, P. R. China\\
  $^3$Georgia Institute of Technology, Atlanta, GA. USA\\
  \small \tt cl122@mail.ustc.edu.cn, jundu@ustc.edu.cn, chl@ece.gatech.edu}
\begin{document}
\ninept
\maketitle
\begin{abstract}
One challenging problem of robust automatic speech recognition (ASR) is how to measure the goodness of a speech enhancement algorithm (SEA) without calculating the word error rate (WER) due to the high costs of manual transcriptions, language modeling and decoding process. Traditional measures like PESQ and STOI for evaluating the speech quality and intelligibility were verified to have relatively low correlations with WER. In this study, a novel acoustics-guided evaluation (AGE) measure is proposed for estimating performance of SEAs for robust ASR. AGE consists of three consecutive steps, namely the low-level representations via the feature extraction, high-level representations via the nonlinear mapping with the acoustic model (AM), and the final AGE calculation between the representations of clean speech and degraded speech. Specifically, state posterior probabilities from neural network based AM are adopted for the high-level representations and the cross-entropy criterion is used to calculate AGE. Experiments demonstrate AGE could yield consistently highest correlations with WER and give the most accurate estimation of ASR performance compared with PESQ, STOI, and acoustic confidence measure using Entropy. Potentially, AGE could be adopted to guide the parameter optimization of deep learning based SEAs to further improve the recognition performance.

\end{abstract}
\begin{keywords}
Acoustic model, state posterior probabilities, cross entropy, speech enhancement, robust speech recognition
\end{keywords}
\section{Introduction}
\label{sec:intro}
Noise robustness is one of the critical issues to make automatic speech recognition (ASR) system widely used in real world today \cite{li2014overview}. A range of approaches has been proposed to tackle the problem to make ASR systems robust against environmental distortions, including front-end processing and robust ASR back-end.
Concerning front-end processing which includes speech enhancement (SE), feature transformation and feature set expansion, its objective is to convert an observed speech signal to a set of input features of the recognition system that are insensitive to environmental distortion while simultaneously containing a sufficient amount of discriminant information \cite{yoshioka2015environmentally}.

As we know, features extracted from clean speech signals contain much more discriminant information than those of corrupted speech. Therefore, if clean speech can be well recovered, better word error rates (WERs) will be obtained. SE is applied to achieve this goal on either signal level or feature level.
Quality evaluation of the resulting enhanced signals is a very complex task that depends on the application field.
For ASR systems, auditory perception is not as important as preserving some acoustic cues that are used by the system to perform the recognition experiments \cite{di2007objective}.
Clearly, one direct way for selecting an optimal SE algorithm for ASR is to perform
recognition experiments. Nevertheless, this approach is time-consuming and requires a large amount of computation and manual transcription costs. An alternative way is to estimate the recognition performance using a distortion value that was originally developed for the objective evaluation of speech performance and represents the difference between degraded speech and its original clean version \cite{di2007objective, yamada2006performance, ling2013performance, fukumori2013estimation, guo2016performance}.
So far, it has been shown that the correlation coefficient between the WER and short-time objective intelligibility (STOI) \cite{taal2011algorithm} is higher than other objective evaluation scores \cite{moore2017speech, thomsen2015speech}.
Besides, \cite{ogawa2016estimating} proposed recognition performance estimation methods based on error type classification without using manually transcribed reference. They can capture the changes in the recognition accuracy caused by the changes in various factors under real conditions and characterize the recognition results.
In \cite{misra2003new, barker1998acoustic}, an acoustic confidence measure usually defined as the entropy of the posterior distribution from the output of the artificial neural network (ANN) was proposed and shown a high degree of correlation with WER, where the discriminatory power of the ANN decreases and the posterior probabilities tend to become more uniform with a higher entropy.

In many cases, SE algorithms for noisy ASR is tuned according to the distortion measures such as perceptual evaluation of speech quality (PESQ) \cite{rix2001perceptual} and STOI.
However, no research has proved that a good value of these distortion measures for enhancement techniques necessarily leads to
a better WER. Actually, some researches have shown that speech enhancement algorithms which achieve a better distortion value may result in a worse ASR performance especially for multi-condition training \cite{chen2018building} because they inevitably introduce distortions that could not be captured by the distortion measures.
There are various factors that affect the recognition accuracy under realistic conditions. It is difficult for distortion measures to capture the changes in the recognition accuracy caused by the changes in any factors other than the one or few factors being focused on \cite{ogawa2016estimating}.
Besides, there are many SE algorithms for noise robust ASR that directly output not the waveform of speech but the speech feature for ASR \cite{bagchi2015combining}. The recognition performance estimation method using the distortion measures which are usually calculated by waveform of the processed speech and clean speech is not easily applicable to these SE algorithms.

\begin{figure*}[!htp]
\centering
\includegraphics[width=0.9\linewidth]{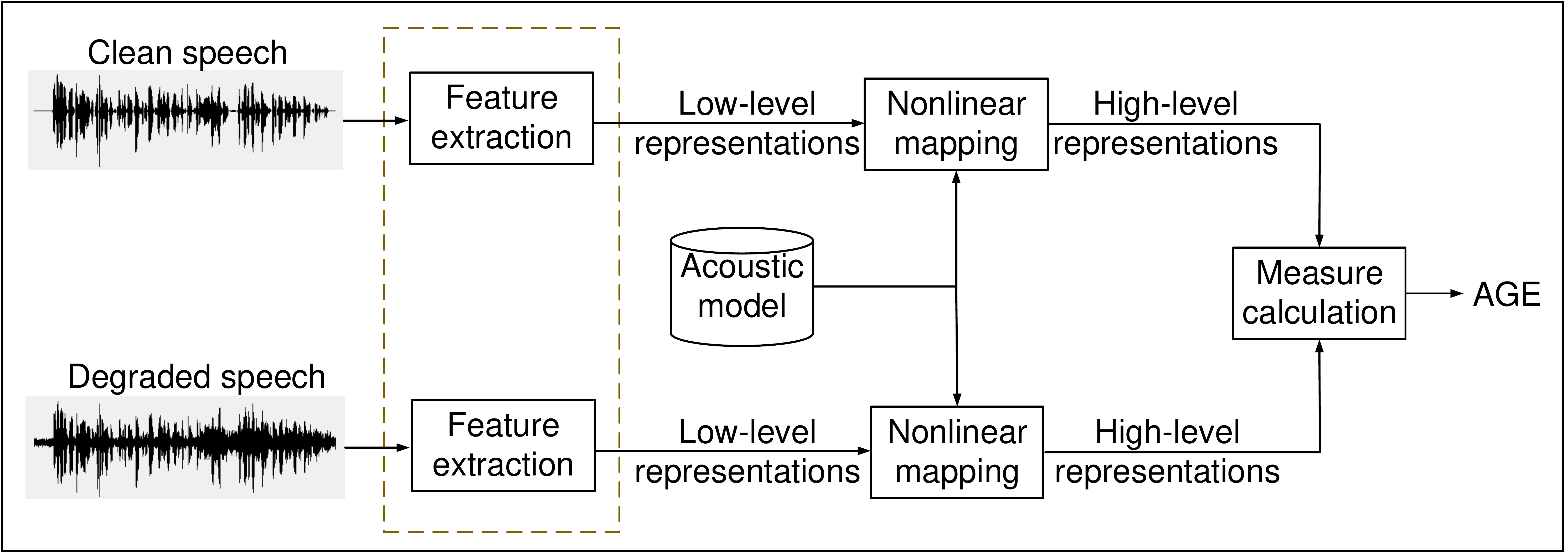}
\caption{ The overall framework of AGE.}
\label{fig_AGE}
\end{figure*}

In this study, we propose an acoustics-guided evaluation (AGE) measure defined as the cross entropy of the state posterior probabilities (SPP) between the degraded speech and the clean speech using ANN-HMM based acoustic model (AM). Experiments demonstrate a consistently highest degree of correlation between AGE and WER compared with the acoustic confidence measure and distortion measures from different aspects including AMs, language models (LMs), SE algorithms, signal-to-noise-ratio (SNR) levels and noise types. Furthermore, AGE gives the most accurate estimation of recognition performance of SE algorithms for noise robust ASR.

\section{AGE}
\label{sec:pagestyle}
The overall framework of AGE is illustrated in Fig.~\ref{fig_AGE}. Generally, it is a measure function of high-level representations (HLR) of both the clean speech and degraded speech calculated by the nonlinear operations between the raw time signals or hand-crafted features and the corresponding AM. The AGE calculation process mainly includes three steps. The first step is the extraction of low-level representations (LLR) which could be from the raw time signals or various hand-crafted features including mel-frequency cepstral coefficients (MFCCs), log-mel-filterbank (FBANK) features and feature-space maximum likelihood linear regression (fMLLR) \cite{povey2011kaldi}.
Due to the lack of ASR acoustic information in LLR, in the second step, the AM which has been well-trained is adopted to map the LLR to HLR which could provide useful acoustic knowledge for better estimation of ASR performance. Specifically, SPP based on ANN-HMM based AM are adopted as the HLR in this study. Therefore, AGE
currently aims to work with the ANN-HMM based ASR system which is also one of the main streams and could be easily applied for other types of ASR systems \cite{yu2017recent}.
Other HLR learned from LLR will be explored for generic AMs in our future work.
The last step is the calculation of AGE which measures the difference between the SPP of the clean speech and degraded speech by a criterion, e.g. cross entropy, Kullback-Leibler divergence and minimum mean squared error (MMSE). Motivated by the training criterion of ANN-HMM based AMs, cross entropy is adopted in this study. Accordingly, AGE is defined as follows:
\begin{equation}
\label{equ_ce}
m=-\frac{1}{N}\sum_{n=1}^{N}\sum_{i=1}^{I}P_C(s_i|\boldsymbol{x}_n^C,\theta)\log P_D(s_i|\boldsymbol{x}_n^D,\theta)
\end{equation}
where $P_C$ and $P_D$ are the SPP of the clean speech and degraded speech, respectively, $I$ is the number of output classes or phonemes, $N$ is the number of frames, $\boldsymbol{x}_n^C$ and $\boldsymbol{x}_n^D$ are the raw time signal or hand-crafted feature vectors of the clean speech and degraded speech for the $n$-th frame, respectively and $\theta$ is the set of parameters of the ANN-HMM based AM.

Distortion measures were originally developed to evaluate the speech performance and not directly correlated with ASR performance, for whom it is difficult to capture various factors that affect the ASR performance. Actually, they only focus on some distortions of the degraded speech. In contrast, AGE considers the acoustic properties by the introduction of HLR, which is directly correlated with ASR system. Furthermore, the difference of the acoustic properties between the clean speech and degraded speech is accurately measured by the cross entropy. AGE can capture the changes of the ASR performance brought by different AMs while distortion measures like PESQ and STOI are invariant to the backend. Besides, distortion measures are not easily applicable to the SE algorithms on feature level, while AGE can be used to evaluate the ASR performance of SE algorithms on both the signal level and feature level.

In comparison to the acoustic confidence measure using the entropy, AGE uses the stereo-data of both clean speech and degraded speech, which potentially gives a more accurate evaluation of the ASR performance. Moreover, AGE is still applicable to real data set where the clean reference signals could be replaced with the signals from close-talking microphones due to the included comparable discriminatory information of them.
Compared with both the acoustic confidence measure and distortion measures, AGE gives the most accurate estimation of ASR performance of SE algorithms, which is shown in Section \ref{222} in detail.
Accordingly, the proposed AGE could be the best choice for the evaluation of the ASR performance of the SE algorithms.

It is worth mentioning that AGE can replace the MMSE as the optimization criterion for ANN based SE algorithms aiming at better recognition performance. Although many advanced objective functions have been investigated recently \cite{chai2018error, fu2018end}. However, these objective functions are not directly correlated with ASR performance. To the best of our knowledge, the optimization criterion directly correlated with ASR performance for ANN based SE algorithms has not yet been studied, we take the lead in proposing a training criterion directly correlated with ASR performance and will disclose more details in our future work due to the space limitation here.
In addition, AGE can replace the acoustic confidence measure to distinguish the ASR performance of each microphone stream for the classifier combination in the context of multi-stream ANN systems. It can also be used to select an optimal AM without increasing the computational resources when there are multiple AMs and their suitable environments are different.

\section{Evaluation procedure}
\subsection{Correlation coefficients}
\label{1}
There are three kinds of commonly used correlation coefficients , namely the Pearson correlation coefficient, the Spearman rank correlation coefficient and the Kendall Tau rank correlation coefficient, while the Pearson correlation coefficient is the most common one \cite{chen2002correlation}. Hence Pearson correlation coefficient is adopted in our study, which is a measure of the linear correlation between two data sets and can be calculated as follows.
\begin{equation}
\label{equ_pcc}
\rho_{xy}=\frac{\sum_{n=1}^{N}(x_n-\overline{x})(y_n-\overline{y})}{\sqrt{\sum_{n=1}^{N}(x_n-\overline{x})^2}\sqrt{\sum_{n=1}^{N}(y_n-\overline{y})^2}}
\end{equation}
This equation can be considered as an expression of a ratio of how much the two data sets $\boldsymbol{x}=[x_1x_2...x_N]^\top$ and $\boldsymbol{y}=[y_1y_2...y_N]^\top$ vary together compared to how much they vary separately \cite{chen2002correlation}.
The magnitude of the correlation coefficient indicates the strength of the correlation and the sign indicates if the correlation is positive or negative.
\subsection{Mapping}
\label{2}
We are interested in measuring the monotonic relation between the AGE and WER. Accordingly, first a mapping is used in order to account for a nonlinear relation between the AGE and WER. The main reason for this mapping procedure is to linearize the data such that we can use the Pearson correlation coefficient. Motivated by \cite{yamada2006performance, thomsen2015speech, taal2011algorithm}, a logistic function is used here:
\begin{equation}
\label{equ_logistic}
f(m)=\frac{100}{1+\exp (am+b)}
\end{equation}
where $a$ and $b$ are constants to be determined by a data-fitting using the least-squares method, $m$ represents the AGE score and $f(m)$ could be considered as an estimator of the WER which is between 0 and 100. Please note that a logistic function is also monotonic and will therefore not influence the monotonicity between the AGE and WER.
Then the performance of AGE is evaluated by means of the Pearson correlation coefficient ($\rho$), which is applied on the mapped objective scores, i.e., $f(m)$. Please note that the same evaluation
procedure is used as with AGE for the acoustic confidence measure and distortion measures.
Since we are only interested in the strength of the correlation, the results of the magnitude of the correlation coefficient denoted as $\rho$ is shown in the following experiments.
\section{Experiments}
\subsection{Experimental setting}
Experiments are conducted on the 1-channel CHiME-4 task \cite{vincent2017analysis}. To make the calculation of the distortion measures (e.g. PESQ and STOI) feasible, we evaluate the correlation of WER with AGE, the acoustic confidence measure and the distortion measures by simulated data that are generated by artificially mixing background noises including cafe (CAF), street junction (STR), public transport (BUS), and pedestrian area (PED) with clean speech data from the development and test data consisting of 410 and 330 utterances respectively.
 The correlation is investigated for the five situations, including different SE algorithms as a preprocessing stage of ASR, different AMs, different LMs, different noise types, and different SNRs.
 The experiments of the top four situations are conducted on the simulated data from the official development and test sets consisting of 1640 and 1320 utterances respectively which has been generated by mixing the abovementioned clean speech data with background noises.
 The experiments of the last situation are conducted on the simulated data constructed by mixing the abovementioned clean speech data with background noises from all the channels at six levels of SNRs (-5dB, 0dB, 5dB, 10dB, 15dB and 20dB) to form 75480 utterances, respectively.
 In this study, the AMs in multi-condition training and clean-condition training modes are adopted, which are from DNN-based official baseline trained by cross entropy minimization and DNNsMBR-based official baseline trained by cross entropy minimization followed by state-level minimum Bayes risk (sMBR) optimization which are both provided by Kaldi \cite{povey2011kaldi} and trained on the fMLLR transformed features of the real and simulated training set from channel 5. Besides, a deep convolutional neural network (DCNN)-based \cite{yu2016deep} AM trained on the FBANK features of the same training set is also adopted. The latter is DNNsMBR-based model and trained on the clean WSJ0 training set.
3-gram, 5-gram and recurrent neural networks (RNN) based official LMs provided by Kaldi are used.

\subsection{Correlation comparison in different situations}
\label{111}
\begin{figure}[!htp]
\centering
\includegraphics[width=0.49\linewidth]{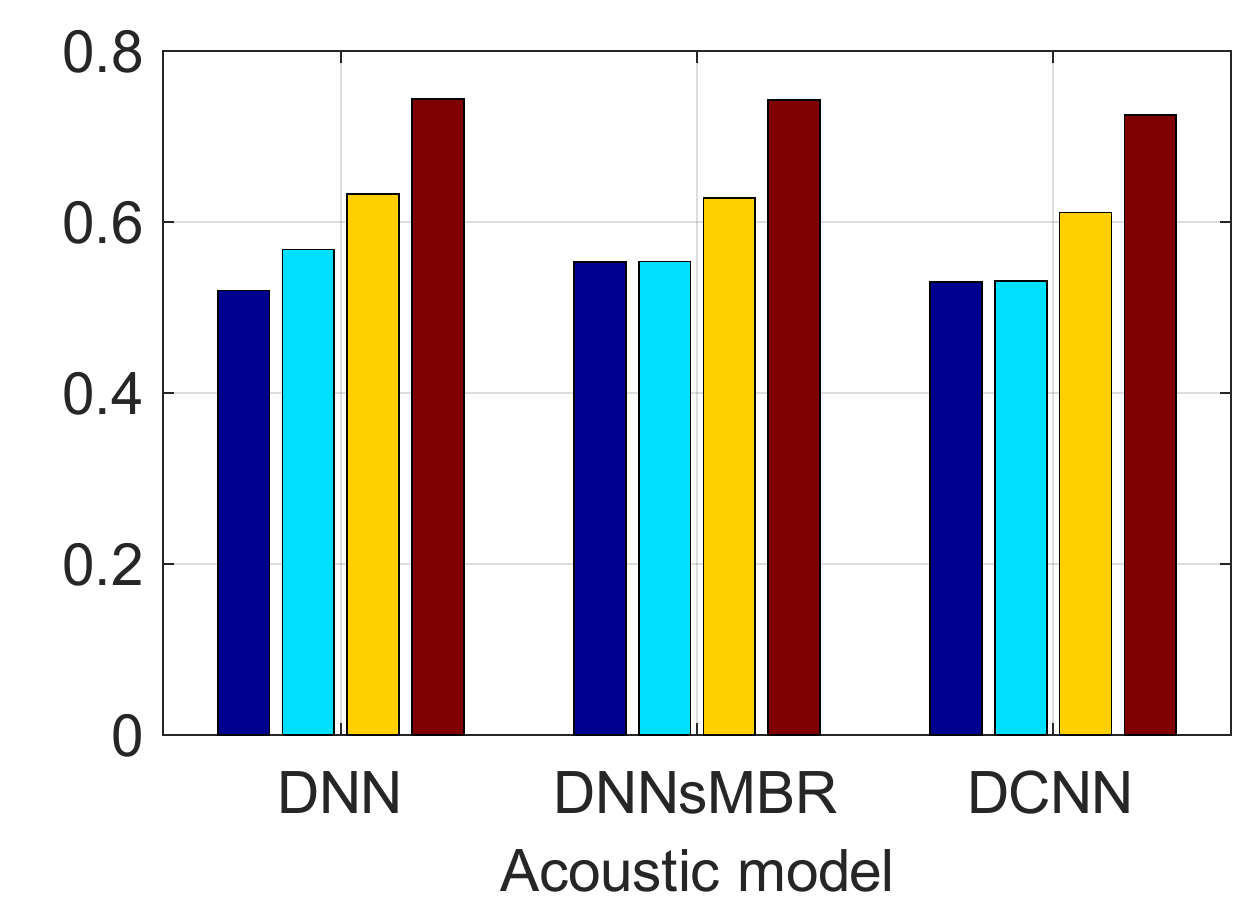}
\includegraphics[width=0.49\linewidth]{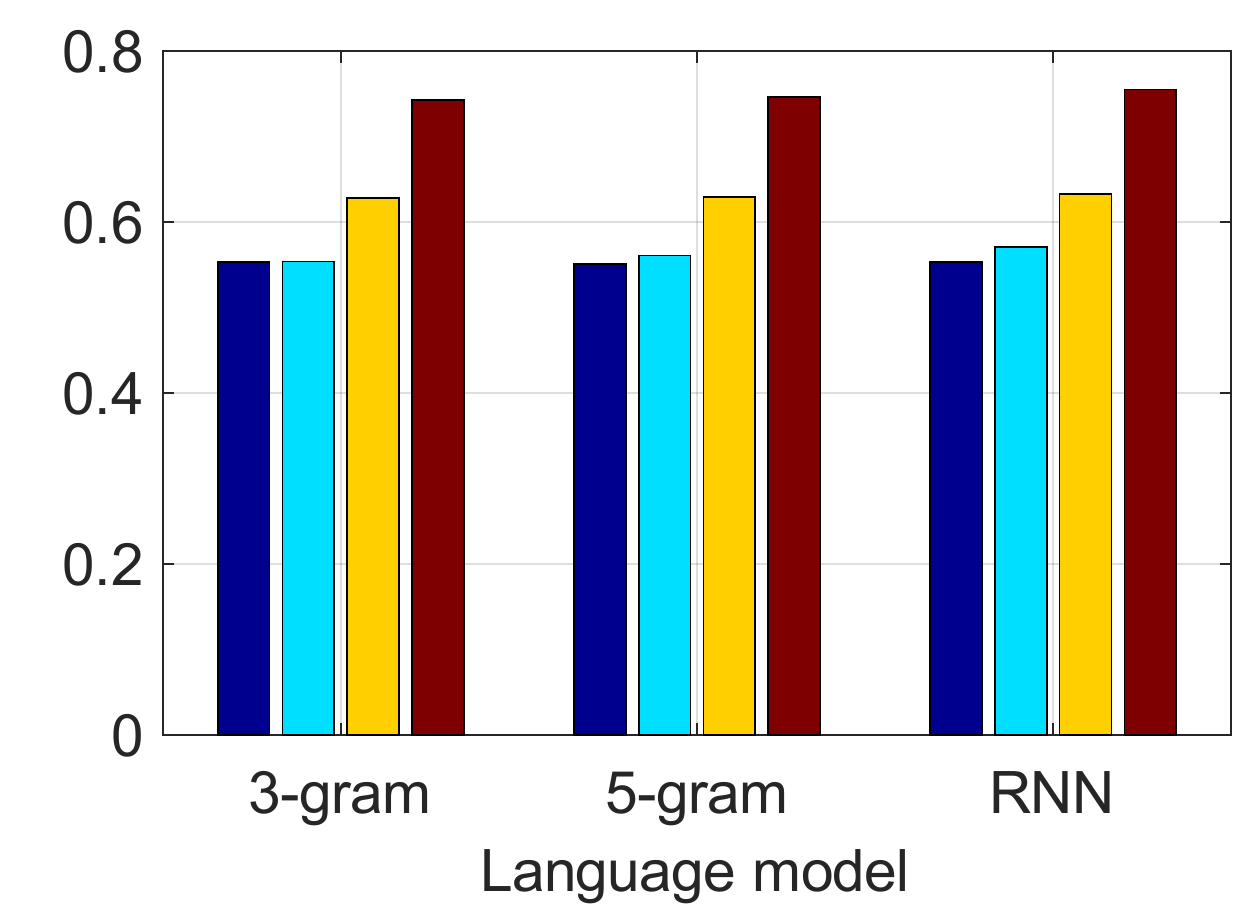}\\
\includegraphics[width=0.35\linewidth]{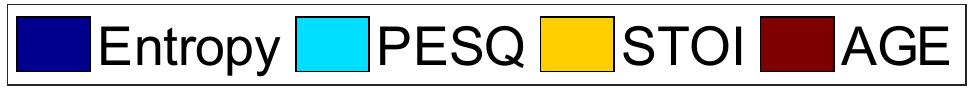}
\caption{Pearson correlation coefficients $\rho$ under different AMs/LMs.}
\label{fig_AM}
\end{figure}

Fig.~\ref{fig_AM} shows a consistently highest degree of correlation between AGE and WER compared with the acoustic confidence measure denoted as Entropy and distortion measures (PESQ and STOI) in AMs using different input features, ANN structures and optimization criterions and LMs including 3-gram, 5-gram and RNN in multi-condition training mode. Moreover, the correlation is robust to both AMs and LMs.
The DNNsMBR-3gram based recognition system was adopted in the following experiments.

\begin{figure}[!htp]
\centering
\subfigure[Multi-condition Training Mode]{
\includegraphics[width=0.249\linewidth]{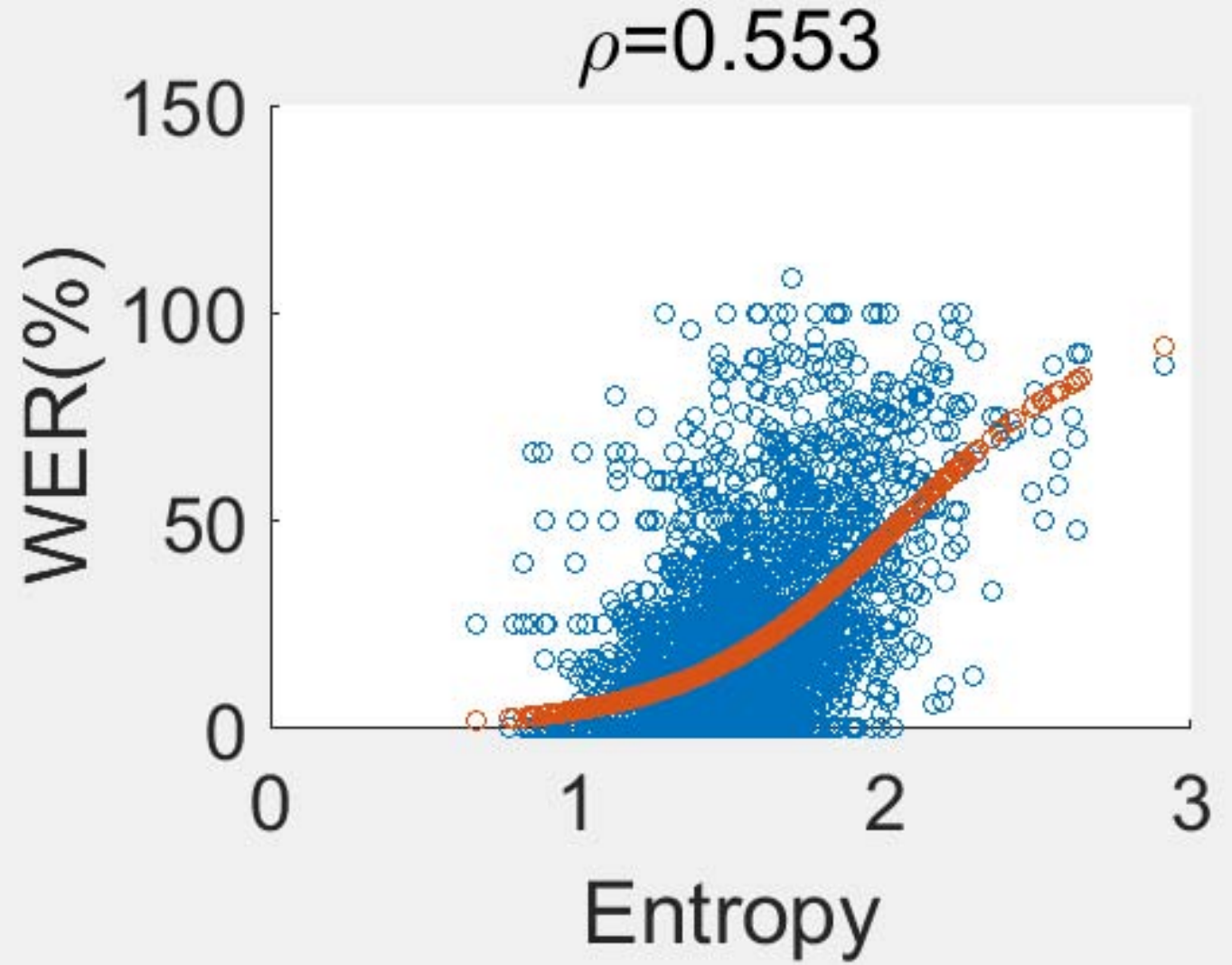}   \includegraphics[width=0.249\linewidth]{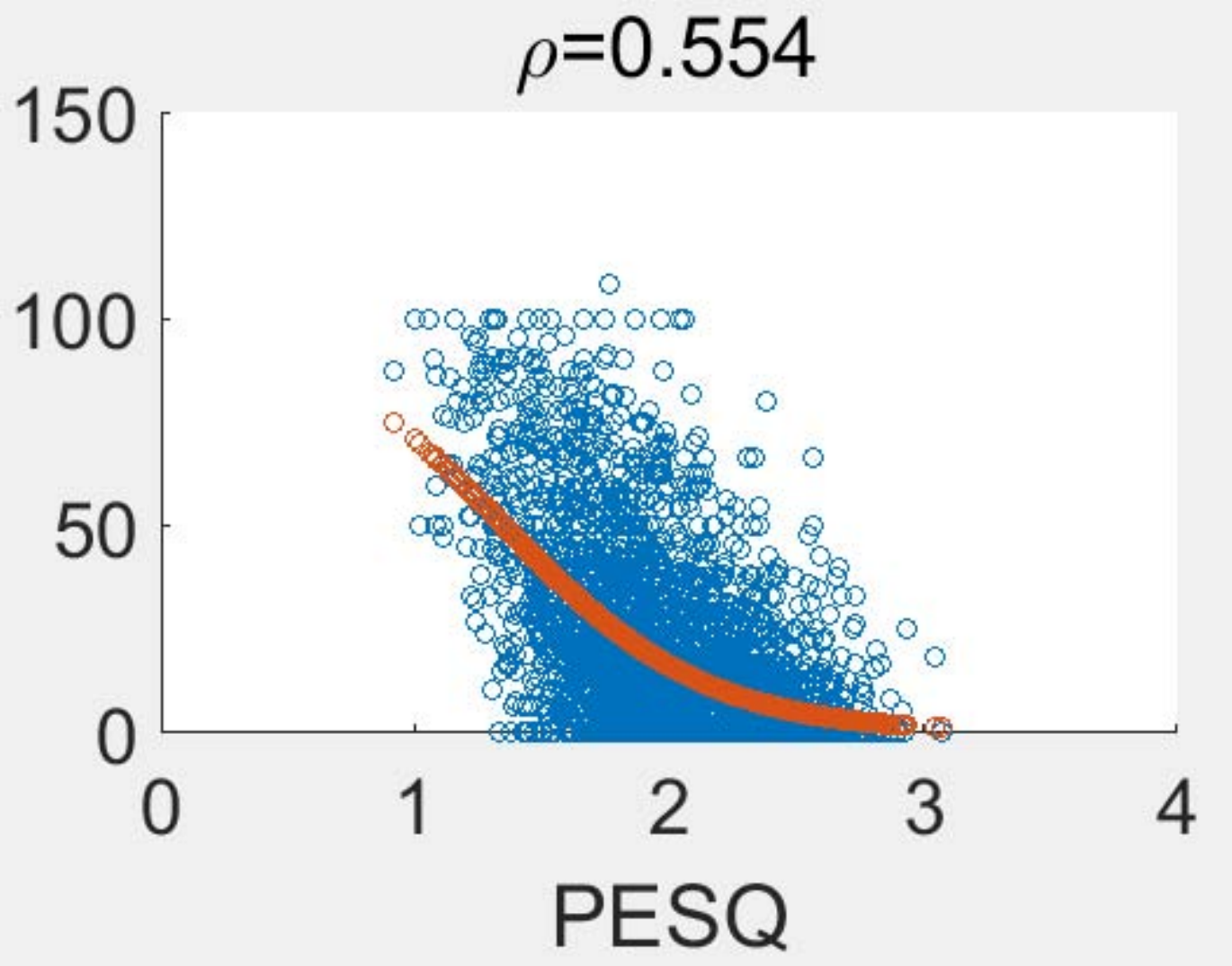}
\includegraphics[width=0.249\linewidth]{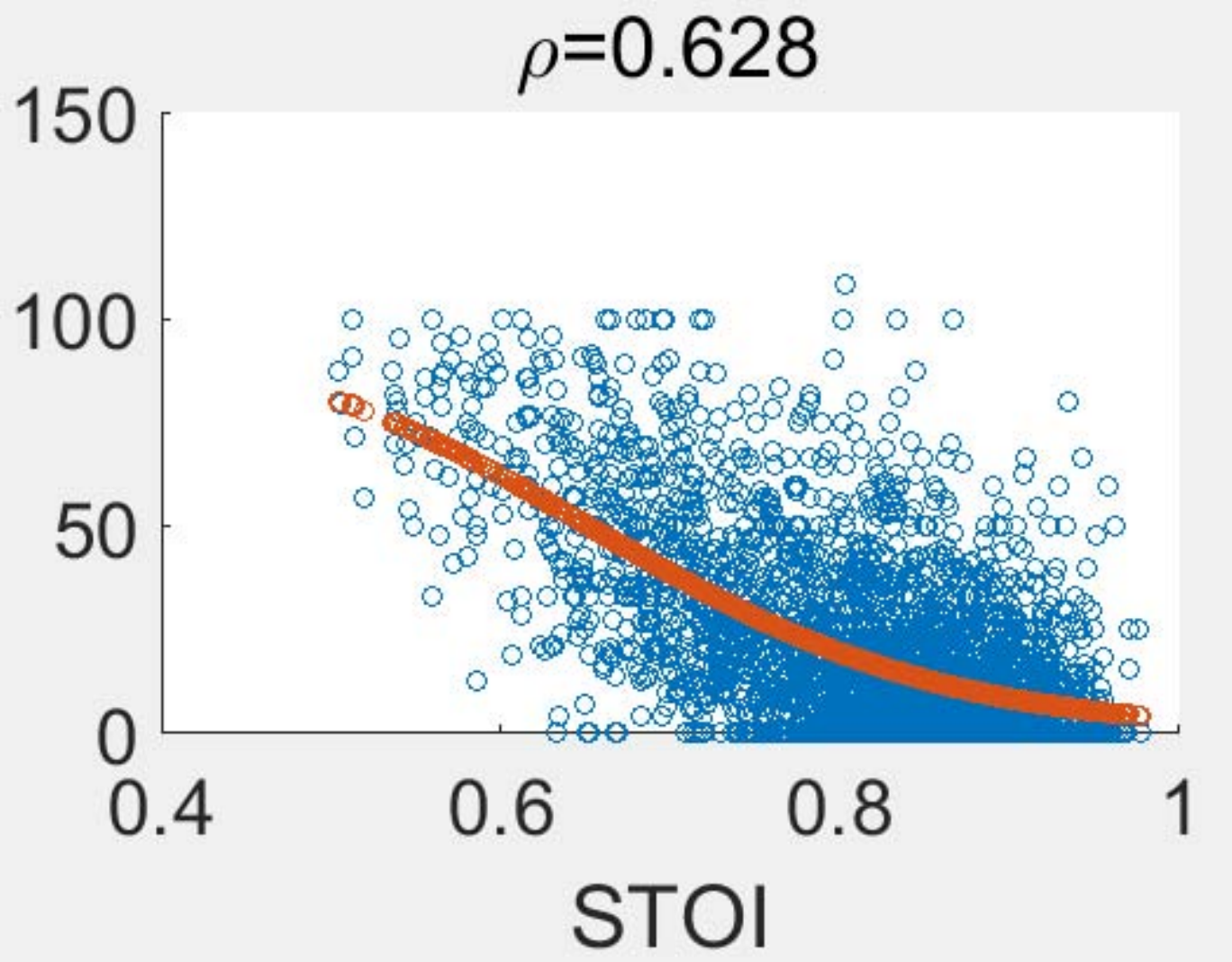}   \includegraphics[width=0.249\linewidth]{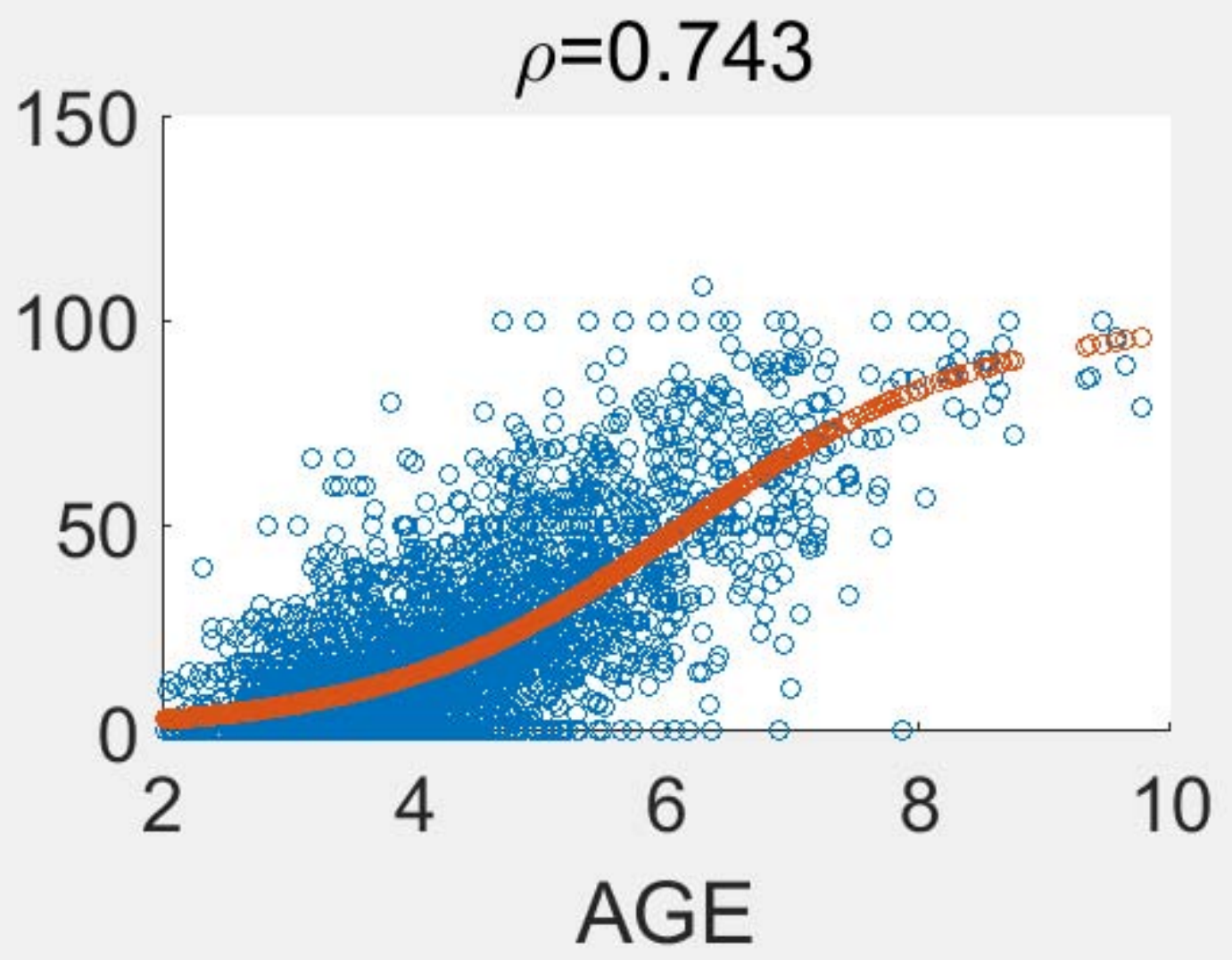}}
\subfigure[Clean-condition Training Mode]{
\includegraphics[width=0.249\linewidth]{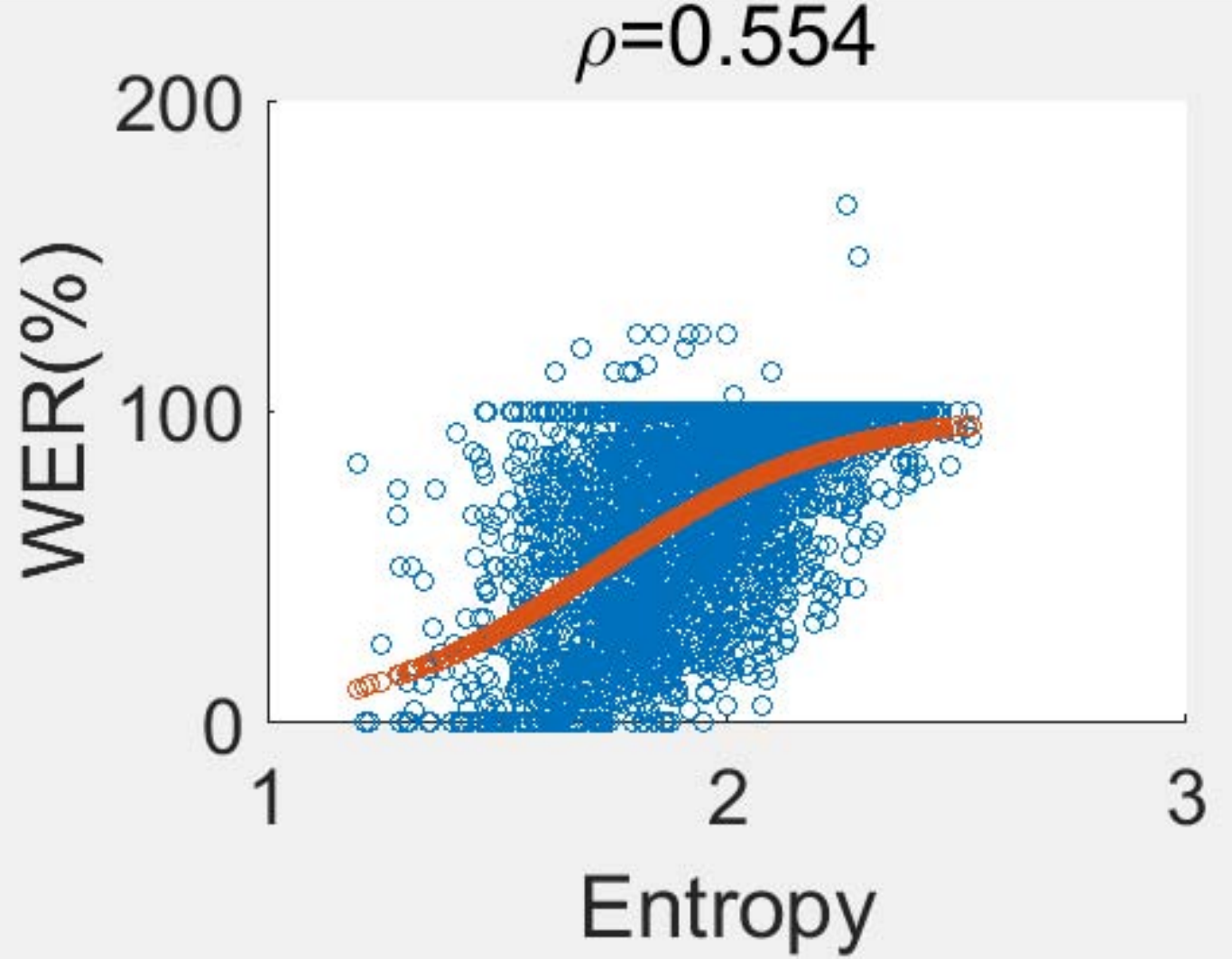}       \includegraphics[width=0.249\linewidth]{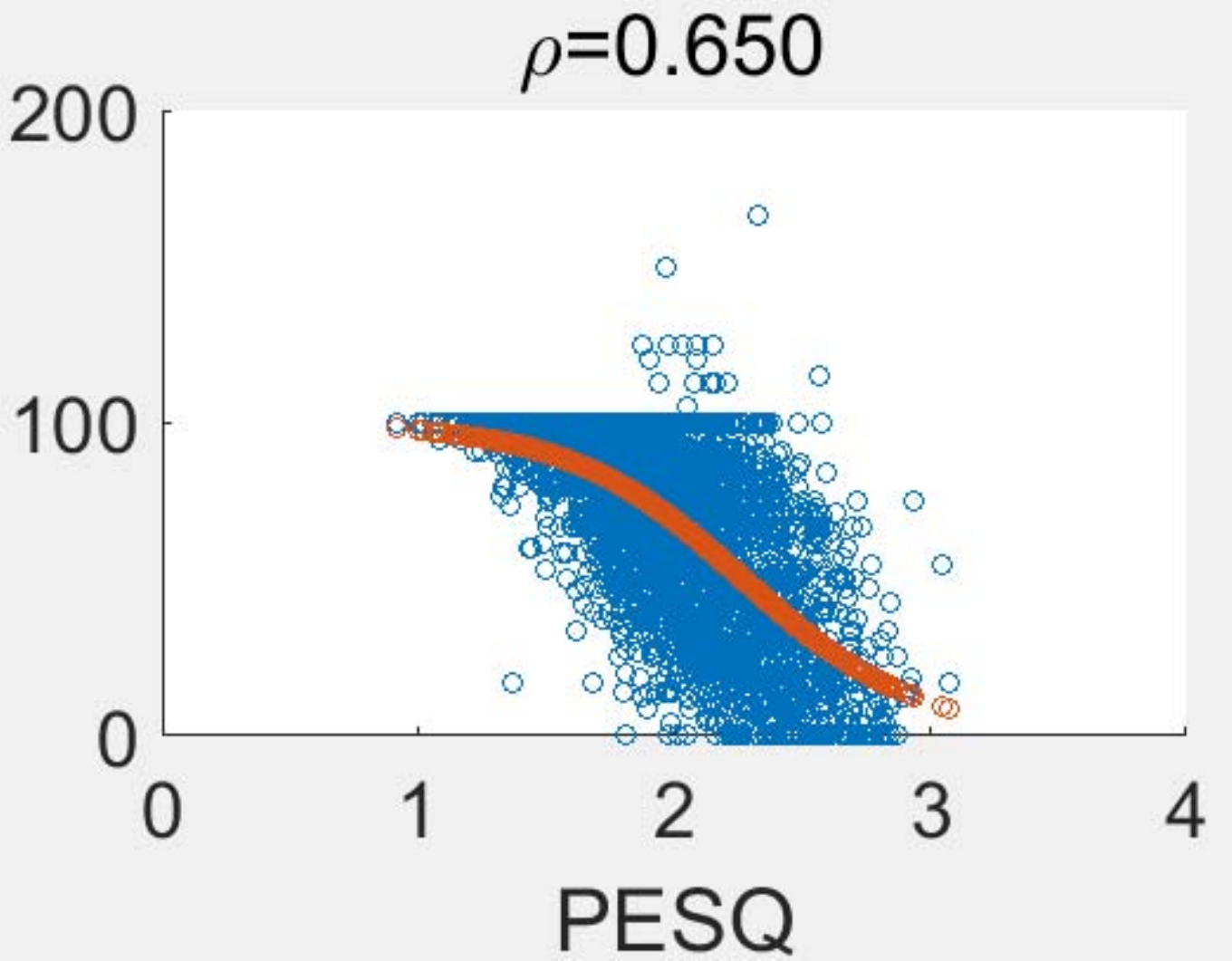}
\includegraphics[width=0.249\linewidth]{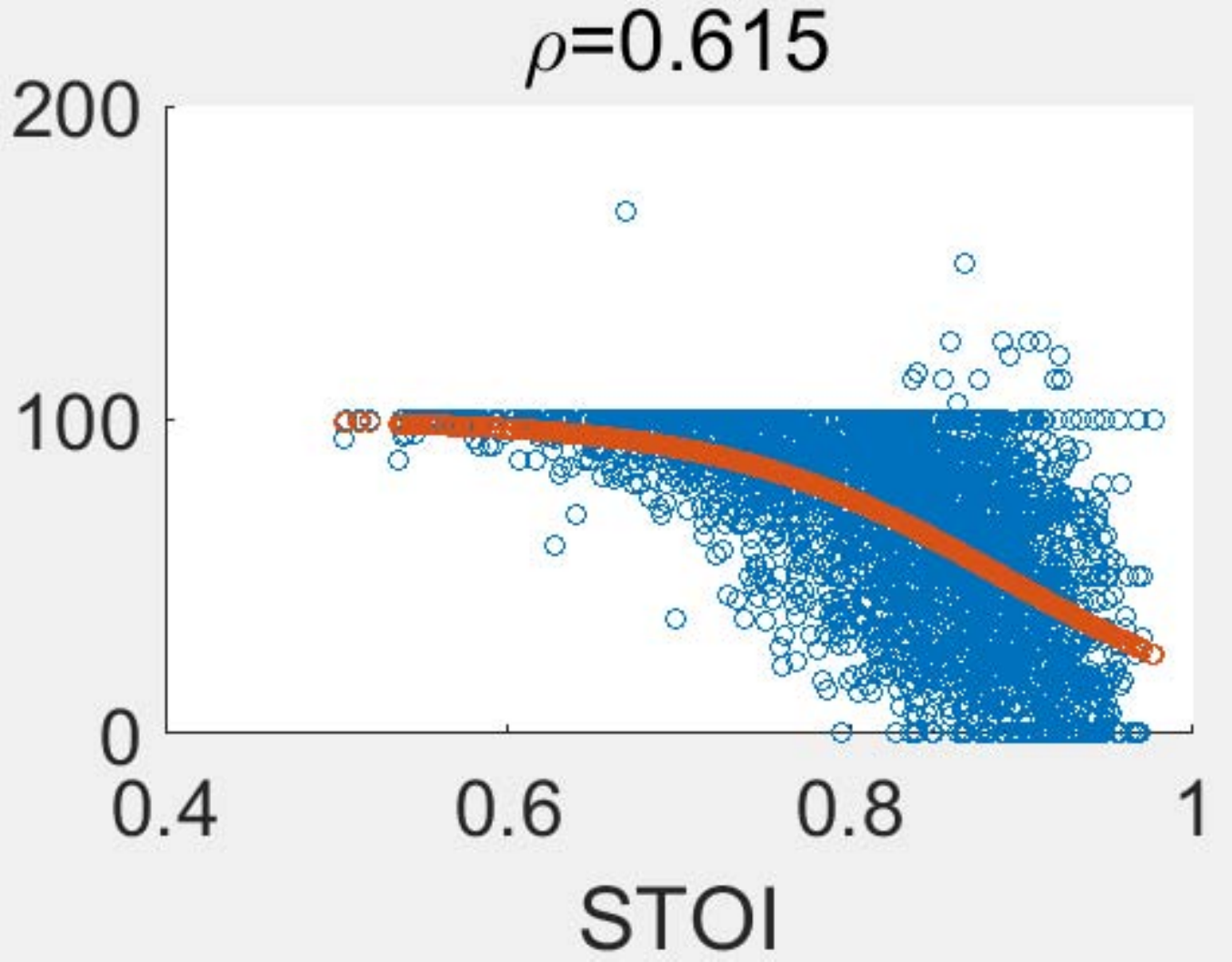}     \includegraphics[width=0.249\linewidth]{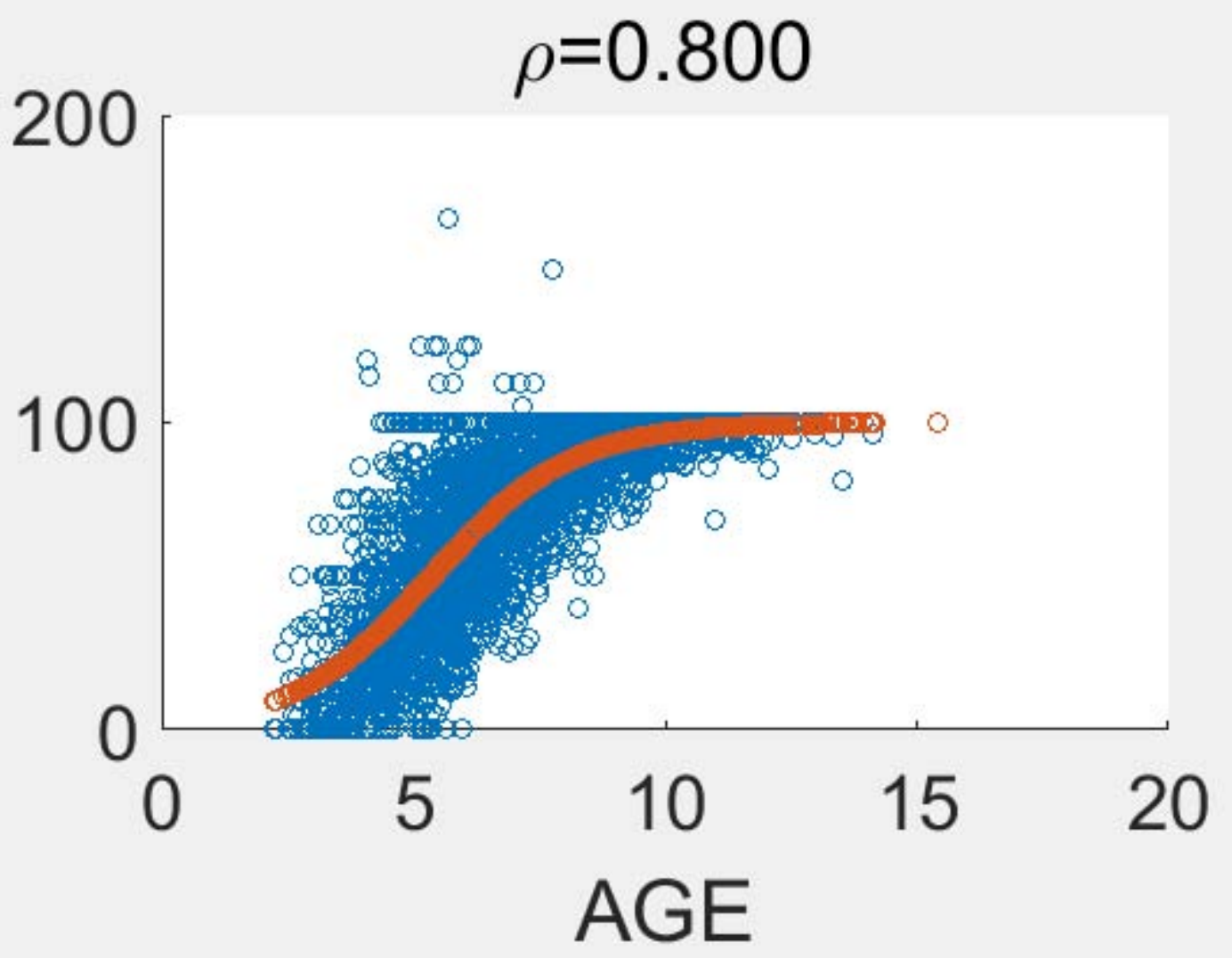}}
 \caption{Scatter plots between the WER and estimation measures.}
\label{fig_mode}
\end{figure}

Fig.~\ref{fig_mode} shows the scatter plots of the relation between WER and the Entropy, PESQ, STOI and AGE, where the red lines represent the applied mapping functions which clearly show good performance by means of a strong monotonic relation with WER. Obviously, AGE has the highest degree of correlation with WER for both clean-condition and multi-condition training. Please note that the correlation score between the AGE and WER for multi-condition training tends to be smaller than the score for clean-condition training because the number of the same scores of WER corresponding to different scores of AGE for multi-condition training is larger, which leads to worse correlation statistics. Unlike the conclusions in \cite{moore2017speech, thomsen2015speech}, we observe that the correlation scores between the STOI and WER tend to be larger than the scores between the PESQ and WER for multi-condition training and the contrary conclusion could be drawn for clean-condition training, which may indicate that the recognition performance depends more largely on speech quality for clean-condition training and on speech intelligibility for multi-condition training.
Besides, it is noted that the acoustic confidence measure and the proposed AGE have positive correlations with WER, on the contrary, PESQ and STOI have negative correlations with WER.

Recognition performance varies with different SE algorithms. Therefore, we investigated the relation with WER by two representative noise reduction algorithms in addition to the reference case of unprocessed noisy speech, namely, an optimally-modified log-spectral amplitude (OM-LSA) speech estimator \cite{cohen2003noise} and a masking-based and DNN enhancement algorithm \cite{narayanan2013ideal}.
Table~\ref{tab_results} shows the consistently strongest correlations between AGE and WER compared with the Entropy, PESQ and STOI in both multi-condition training and clean-condition training modes for different enhancement algorithms.
\begin{table} [!htp]
\newcommand{\tabincell}[2]{\begin{tabular}{@{}#1@{}}#2\end{tabular}}
\caption{\label{tab_results} {Pearson correlation coefficients $\rho$ for different SE algorithms.}}
\vspace{0.1mm}
\centerline{
\begin{tabular}{|@{}p{2.2cm}<{\centering}@{}|@{}p{1.3cm}<{\centering}@{}|@{}p{1.2cm}<{\centering}@{}|@{}p{1.2cm}<{\centering}@{}|@{}p{1.2cm}<{\centering}@{}|@{}p{1.2cm}<{\centering}@{}|}
\hline
\multirow{1}{*}{\tabincell{c}{Training mode}}&\multirow{1}{*}{\tabincell{c}{}}&\multirow{1}{*}{\tabincell{c}{Entropy}}&\multirow{1}{*}{\tabincell{c}{PESQ}} &\multirow{1}{*}{\tabincell{c}{STOI}} &\multirow{1}{*}{\tabincell{c}{AGE}} \\
\hline
\multirow{3}{*} {Multi-condition}  &Noisy   &0.553 &	0.554  &	0.628&0.743\\
&OM-LSA &	0.589 &	0.582& 	0.624&0.726\\
&DNN &0.676 &0.600 	&0.639&0.744\\
\hline
\multirow{3}{*} {Clean-condition}  &Noisy   &0.554 &0.650  &	0.615&0.800\\
&OM-LSA &	0.689 &	0.691& 	0.645&0.794\\
&DNN &0.670 &	0.664 	&0.625&0.778\\
\hline
\end{tabular}}
\end{table}

\begin{figure}[!htp]
\centering
\includegraphics[width=0.49\linewidth]{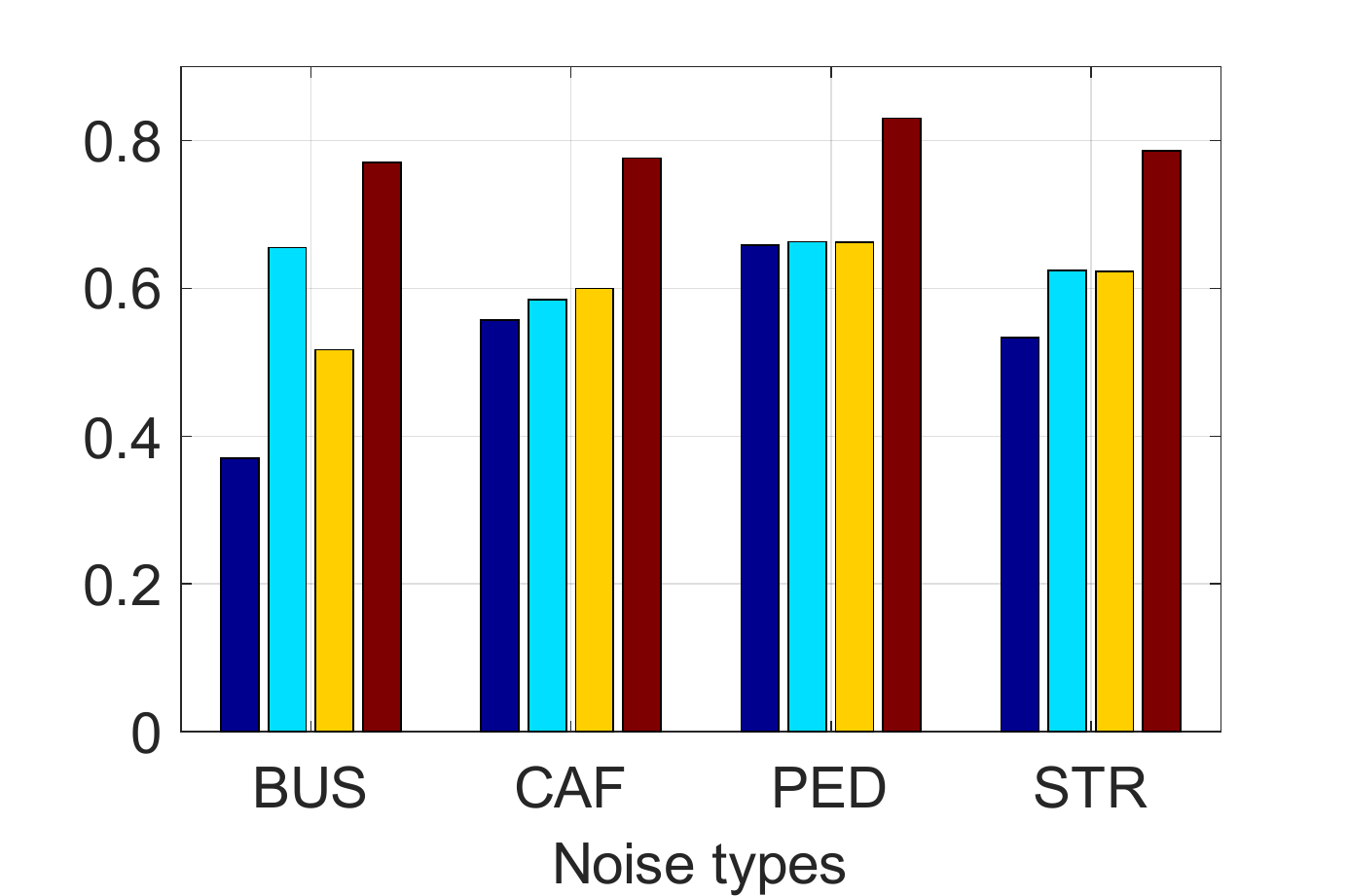}
\includegraphics[width=0.49\linewidth]{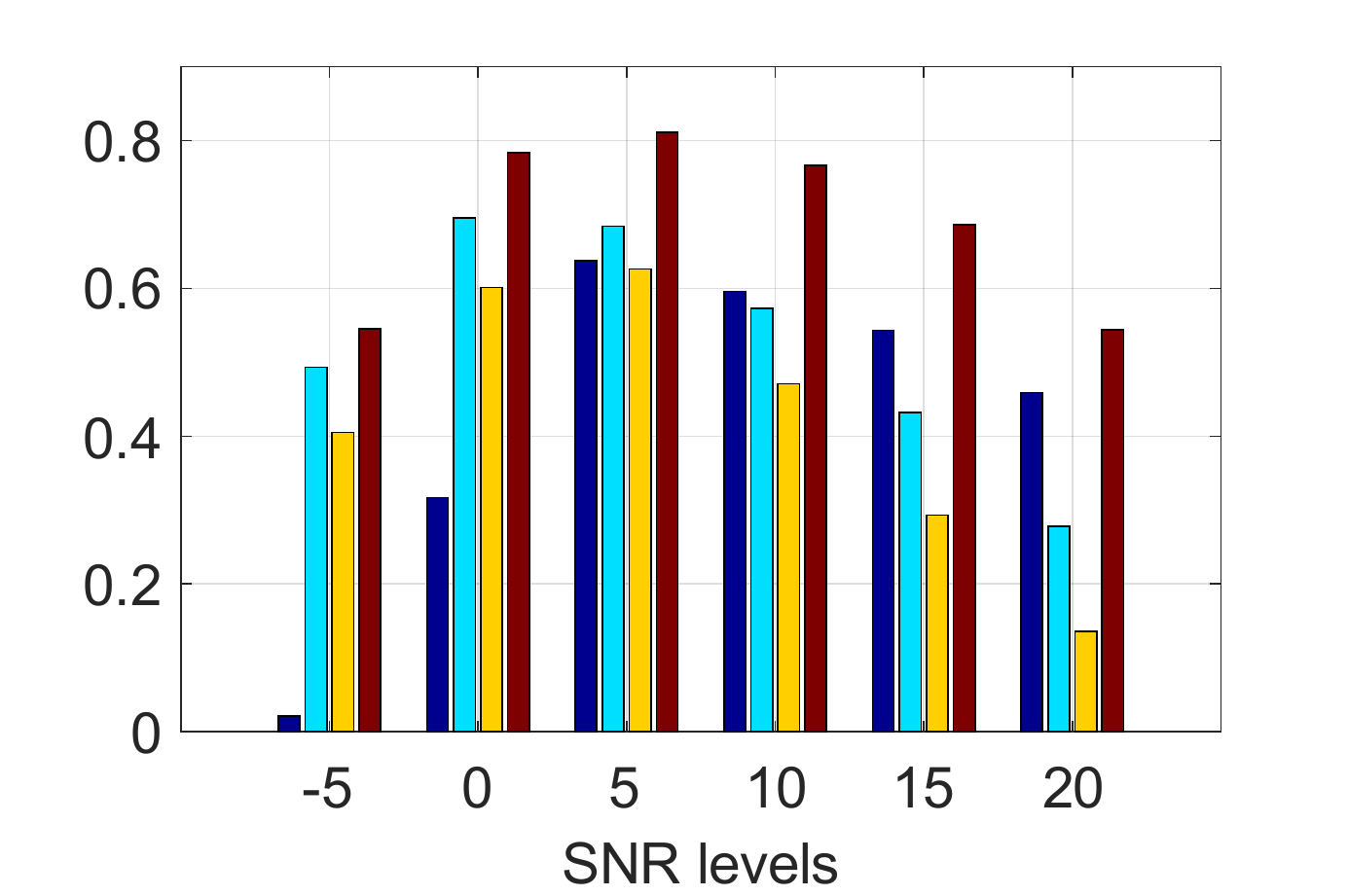}\\
\includegraphics[width=0.35\linewidth]{1legend.pdf}
\caption{Pearson correlation coefficients $\rho$ in clean-condition training mode for different noise types and SNR levels.}
\label{fig_SNRandNoise}
\end{figure}

The relationship is also investigated for different noise conditions and different SNR levels, where AGE still consistently shows strongest monotonic relationship to the WER compared with Entropy, PESQ and STOI shown in Fig.~\ref{fig_SNRandNoise}. The correlation becomes weak in very high SNR levels where most values of WERs are 0\% or in very low SNR levels where most values of WERs are 100\% due to the negative influence of these points which have the same WERs but different values of the evaluation measures.

\subsection{Comparison of evaluation accuracy}
From Table~\ref{tab_results2}, we can see that the multi-condition training mode could bring better recognition performance compared with the clean-condition training mode. However, the improvements could not be evaluated by distortion measures due to their invariance to the backend. In contrast, the acoustic confidence measure and AGE both related to the backend could evaluate it, where the smaller values of them are, the lower WER is.
Besides, there are many cases where the recognition performance could not be accurately evaluated by both the distortion measures and acoustic confidence measure, e.g., the smaller Entropy of OM-LSA does not bring a decline in WER compared with that of DNN in both multi-condition training and clean-condition training modes, better PESQ or STOI of enhanced speech does not bring improvements of recognition performance in multi-condition training mode, the same Entropy of DNN as that of the unprocessed noisy speech does not bring the same WER, and worse STOI of OM-LSA brings a decline in WER for clean-condition training.
\label{222}
\begin{table} [!htp]
\newcommand{\tabincell}[2]{\begin{tabular}{@{}#1@{}}#2\end{tabular}}
\caption{\label{tab_results2} {The average evaluation measure scores and WERs of different SE algorithms for ASR on the official simulated development and test sets (2960 utterances).}}
\vspace{0.1mm}
\centerline{
\begin{tabular}{|@{}p{2.3cm}<{\centering}@{}|@{}p{1.4cm}<{\centering}@{}|@{}p{1.4cm}<{\centering}@{}|@{}p{1.4cm}<{\centering}@{}|@{}p{1.4cm}<{\centering}@{}|@{}p{1.4cm}<{\centering}@{}|}
\hline
\multirow{1}{*}{\tabincell{c}{Training mode}}&\multirow{1}{*}{\tabincell{c}{}}&\multirow{1}{*}{\tabincell{c}{Noisy}} &\multirow{1}{*}{\tabincell{c}{OM-LSA}} &\multirow{1}{*}{\tabincell{c}{DNN}} \\
\hline
\multirow{5}{*} {Multi-condition}  &Entropy   &1.52 &	1.65  &	1.71\\
&PESQ &	2.00 &	2.25& 	2.30\\
&STOI &0.819 &0.808 	&0.846\\
&AGE &4.23	&4.45	&4.39\\
&WER(\%) &19.39	&25.70	& 24.46 \\
\hline
\multirow{5}{*} {Clean-condition}  &Entropy   &1.92 &	1.89  &	1.92\\
&PESQ &	2.00 &	2.25& 	2.30\\
&STOI &0.819 &	0.808 	&0.846\\
&AGE &6.81	&5.99	&5.68\\
&WER(\%) &65.52	&57.4	&51.59  \\
\hline
\end{tabular}}
\end{table}
In contrast, AGE could accurately evaluate the recognition performance of different SE algorithms for both multi-condition training and clean-condition training.
Please note that both the ASR and SE performance of the SE algorithms which were not originally designed for the improvement of STOI (e.g. OM-LSA) or other distortion metrics could not be well evaluated by corresponding distortion measures.

\section{Conclusion}
In this study, we propose a measure, i.e., AGE to evaluate the performance of the SE algorithms for noise-robust ASR without using reference transcriptions, LMs and recognition process. Compared with both the acoustic confidence measure and distortion measures, AGE shows the highest correlation with WER and gives the most accurate estimation of ASR performance. Moreover, AGE could be directly adopted as the optimization criterion of the ANN based SE algorithms for ASR instead of the conventional MMSE criterion, which will be explored in our another work.


\newpage
\bibliographystyle{IEEEbib}
\bibliography{strings,refs}

\end{document}